# On the common solution within a framework of single matrix algorithm for radical suppression of background products in heavy-ion induced nuclear reaction

Yu.S.Tsyganov

**Abstract**

Application of real-time matrix algorithm in heavy ion induced full fusion nuclear reactions of super heavy elements synthesis is reviewed in brief. An extended algorithm, for the case of the recoil detection efficiency is not close to 100% has been proposed.

1**. Introduction**

Recently, at the Dubna Gas-filled Recoil Separator (DGFRS, [1]) more than 17 new nuclides with Z=110 to 118 have been synthesized [2]. It should be noted that some of these experimental results were clearly confirmed by the independent experiments [3-4], involving studying of chemical properties the synthesized atoms. To concentrate the reader's attention three major components of the success may be listed, namely:

• electromagnetic recoil separator design must provide not only acceptable value of the nuclide transportation efficiency (tens of percent), but also a significant suppression of background products;

• heavy-ion beam intensity should be high enough to provide nuclide under investigation formation;

• detection system must be not only quite informative to provide identification of the nuclide, but it should contribute to the process of deep suppression of backgrounds products .

Following the above points, the author formulated that an algorithm design, the goal of the present paper, is in fact can be considered as a supplement to the third point.

## 2. Real-time detection mode for radical suppression of the backgrounds in experiments aimed to the synthesis of super heavy elements (SHE).

Usually, to reach high total SHE experiment efficiency, one use extremely high (n*$10^{12}$ to $10^{13}$ pps, n > 1) heavy ion beam intensities. It means, that not only irradiated target, sometimes (frequently) made on highly radioactive actinide material, should not be destroyed during long term experiment, but the in-flight recoil separator and its detection system should provide backgrounds suppression in order to extract one-two events from the whole data flow. Typically, the DGFRS provides suppression of the beam-like and target-like backgrounds by the factors of[1] ~$10^{15}$-$10^{17}$ and $10^4$-5*$10^4$, respectively. Nevertheless, under real circumstances, total counting rate above approximately one MeV threshold is about tens to one-three hundreds[2] events per second. Therefore, during, for example one month of irradiation about $30*10^5*100 = 3e+08$ multi - parameter events are written to the hard disk during a typical SHE experiment at the DGFRS.

---

[1] Depending on the reaction asymmetry ( projectile to target mass ratio)

To avoid a scenario that result of the SHE experiment (one-two-three decay chains per month) can be represented as a set of random signals real-time search technique to suppress the probability for detected event to be a random has been designed and successfully applied.

Note, that in the reactions with $^{48}$Ca as a projectile, the efficiency of SHE recoiling products detection both by silicon and TOF detector is close to 100%. Namely recoil-first (second) correlated alpha decay signal was used as a triggering signal to switch off the cyclotron beam for a definite (seconds-minutes) and, therefore, detection of forthcoming alpha decays were in fact "background-free". The basic idea to apply such detection mode is to transform the main data flow to the discrete form [5]. To demonstrate this technique successful application in the Fig.1 the spectrum of $^{287}$114 nuclei is shown [6]. Total α-particle-like energy spectra of events that stopped the beam during $^{242}$Pu+$^{48}$Ca experiment. The solid histogram shows the energies of events that switched off the beam that were followed by a beam-off SF event or beam-off α-particles with $E_α$ = 8.9 -10 MeV and a beam-on SF event, with a total elapsed time interval up to 14 min.

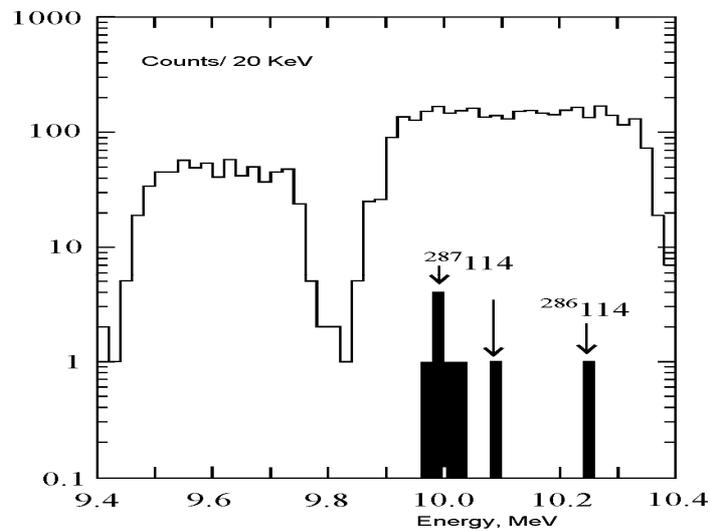

**Fig.1** Spectrum of Z=114 (A=287) detected alpha decays.

Note, that background suppression factor estimates as $\sim \frac{C_n^m}{K}$ for the parameter of probability of the whole event to be a random [7]. Here n – the whole number of alpha decays, m- number of alpha decay, which stops the beam and K – suppression factor for single event($\sim 10^2$-$10^3$). It means, for instance, in the case of four alpha decay chains event, additional suppression factor is about $\sim 10^9$ if first alpha particle signal provides beam switching. Schematic of the whole

---
[2] Including events having only TOF signal and zero energy ( below energy threshold)

process creating a background-free detection of decays is shown in the Fig.2. It takes about 175 μs to provide full beam switching after detection a correlation recoil-alpha sequence including additional time delay (~ 60 μs) related with cyclotron operation.

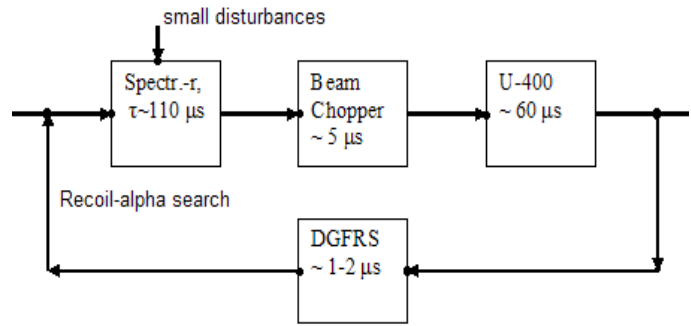

**Fig.2** Flow chart of the whole real-time process which is applied for radical suppression of background products in SHE element synthesis experiments. Delay times are shown for each component.

3. **Extended algorithm**

Unfortunately, in contrast to relatively high projectile, like $^{40}$Ar and $^{48}$Ca, when using projectiles from oxygen to magnesium, real threshold of the detection system sometimes does not allow recoil detection with close to 100% efficiency (see Fig.3).

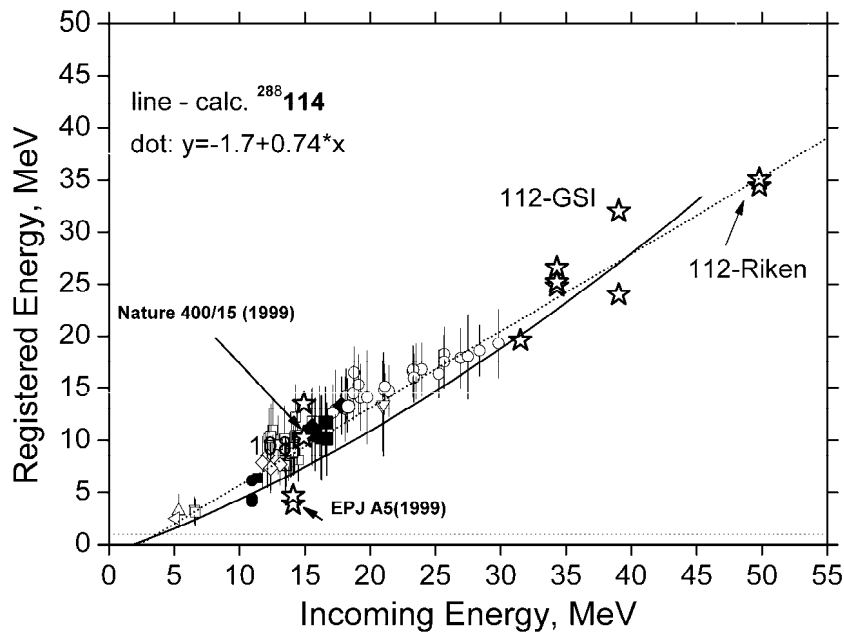

**Fig.3** The dependence of registered heavy recoil energies against calculated incoming ones for the DGFRS detecting module. Line corresponds to calculations [8] for $^{288}$114 recoil (0 < $E_{in}$ < 40 MeV ). Six events of Z=112 element synthesized at GSI [3,9] are shown by stars in the right-upper corner. Two events of Z=112 were synthesized in Riken [10] are shown by stars too. Lower dot line shows energy threshold of about 1 Mev.

It seems more realistic in this case to use both recoil and alpha signals as a first chain of multi - chain event to provide beam switching. Of course, one assume, that the mentioned above value of the recoil detector efficiency is compatible with alpha- particle one. In the second, this value, in contrast to the one, corresponding to reactions with $^{48}$Ca projectile, may be varied with varying of threshold level during an experiment. The schematic of such algorithm is represented in the Fig.4.

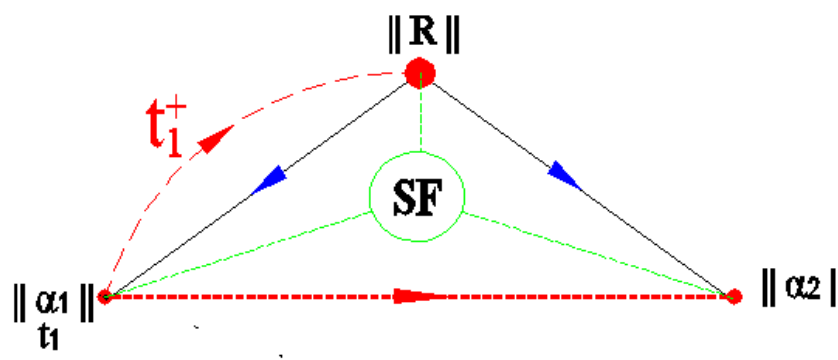

**Fig.4** The schematic of the extended algorithm of real-time search of correlated sequences. In the case of no beam stop is generated by first recoil-alpha sequence elapsed time of this alpha particle event is written to the appropriate matrix cell of the "recoil" matrix. In principle the same operation can also be provided with the second alpha decay event.

Loop $t_1^+$ denotes that in the case of "no success"[3], that is, if no beam stop is generated after recoil-alpha sequence detection, then the elapsed time of coming (and detected) alpha particles is written into the appropriate cell of the first ("recoil") matrix. Hence, the second alpha particle in the multi chain decay event will provide beam switching if time interval between two alpha decays is less than pre-setting resolved time for first recoil-alpha correlation. In the case of one want to use separate pre-settings, it is necessary to use matrix cell in the form of two[4]-dimensional structure, not only write the elapsed time of the event, but also use particle identifier.

Of course, as a partial drawback of such approach one should mention, that the reported solution can be considered as "common" if all the decay life-times are of the same order. Otherwise one should use maximum preset time, and therefore, create extra break points during the process of target irradiation by intense heavy ion beam.

## 4. Conclusion

With a short reviewing of real-time algorithm application aimed to the radical suppression of backgrounds in heavy ion induced full fusion nuclear reaction, the most common solution based on extended algorithm version is suggested. It is estimated, that because of realistic spectrometer energy thresholds, such form of algorithm will be definitely useful in the case when the efficiency of recoil detection is far from 100%.

---

[3] No beam interrupt

[4] Or even more than two. From the viewpoint of practical realization author use C++ code and one can easily use "structure" or "class" object. To a first approximation, second component may be as an assuming number of decay chain in a multi- decay event, of course, starting from a recoil signal.

## 5. Acknowledgements

Author is grateful to Drs. Utyonkov, Polyakov and Shirokovsky for their continuous support. This paper is supported in part by RFBR Grant №07-02-0029.

**References**

[1] K.Subotic et al., // Nucl. Instrum. and Meth. In Phys. Res. A . 2002. V.481 p.71-80

[2] Yu.Ts.Oganessian et al. // Phys. Rev. C. 2006.  v.74. 044602

[3] S.Hofmann et al. //To be printed in Eur.Phys. J.

[4] R.Eihler et al. // Nature, Vol.447/3 (2007) 72-75

[5] Yu.S.Tsyganov et al. // Nucl. Instrum. and Meth. In Phys.Res.  A513(2003) 413-416

[6] Yu.Ts.Oganessian et al. // Phys.Rev. C 70 064609(2004)

[7] Yu.S. Tsyganov // Nucl. Instrum. and meth. In Phys. Res. 2007. A573. p.161-164

[8] Yu.S.Tsyganov // JINR communication E13-2006-77. Dubna, 2007.

[9] S.Hofmann et al. // Eur. Phys. J. A 354 (2002) 147

[10] K.Morita et al. // J. of the Phys. Soc. of Japan Vol. 76, No.4 (2007) 043201